\newif\ifpreprint
\preprinttrue
\newif\ifcomment
\newif\ifnprl
\newif\ifdraft
\ifpreprint
\documentclass[ALICE,manyauthors]{cernphprep}
\usepackage[comma,square,numbers,sort&compress]{natbib}
\else
\documentclass[10pt,aps,prl,twocolumn,superscriptaddress,altaffilletter,nobibnotes,nofootinbib]{revtex4-1}
\usepackage{graphicx}
\fi 
\usepackage[utf8]{inputenc}
\usepackage[T1]{fontenc}
\usepackage{hyperref}
\usepackage{lineno}
\usepackage{epsfig}
\usepackage{color}
\definecolor{dgreen}{cmyk}{1.,0.,1.,0.1}      % dark green
\definecolor{orange}{cmyk}{0.,0.353,1.,0.}    % orange

\newcommand{\com}[1]    {\relax}

\def\dvers {v8}

\def\PbPb  {\mbox{Pb--Pb}}

\def\snn   {\mbox{$\sqrt{s_{_{\rm NN}}}$}}
\def\pt    {p_{\rm{T}}}

\graphicspath{{.img/}}

\begin{document}%
%%%%%%%%%%%%%%%  Title page %%%%%%%%%%%%%%%%%%%%%%%%
\ifpreprint
\PHyear{2021}
\PHnumber{031}      % required, will be obtained from PH
\PHdate{19 February}  % required, will be obtained from PH
\begin{titlepage}
\fi
\title{Measurements of mixed harmonic cumulants in Pb--Pb collisions at $\mathbf{\sqrt{{\textit s}_{\rm NN}}}=5.02$~TeV}
\ifpreprint
\ShortTitle{Mixed Harmonic Cumulants in Pb--Pb collisions}   % appears on right page headers
\Collaboration{ALICE Collaboration\thanks{See Appendix~\ref{app:collab} for the list of collaboration members}}
\ShortAuthor{ALICE Collaboration} % appears on left page headers, do not change
\fi
\ifdraft
\begin{center}
\today\\ \color{red}DRAFT \dvers\ \hspace{0.3cm} \$Revision: 5573$\color{white}:$\$\color{black}\vspace{0.3cm}
\end{center}

\fi
\begin{abstract}
Correlations between moments of different flow coefficients are measured in Pb--Pb collisions at $\sqrt{s_{\rm NN}} = 5.02$~TeV recorded with the ALICE detector. These new measurements are based on multiparticle mixed harmonic cumulants calculated using charged particles in the pseudorapidity region $|\eta|<0.8$ with the transverse momentum range $0.2 < p_{\rm T} < 5.0$ GeV/$c$. The centrality dependence of correlations between two flow coefficients as well as the correlations between three flow coefficients, both in terms of their second moments, are shown. In addition, a collection of mixed harmonic cumulants involving higher moments of $v_2$ and $v_3$ is measured for the first time, where the characteristic signature of negative, positive and negative signs of four-, six- and eight-particle cumulants are observed, respectively. The measurements are compared to the hydrodynamic calculations using iEBE-VISHNU with AMPT and TRENTo initial conditions. It is shown that the measurements carried out using the LHC Run 2 data in 2015 have the precision to explore the details of initial-state fluctuations and probe the nonlinear hydrodynamic response of $v_2$ and $v_3$ to their corresponding initial anisotropy coefficients $\varepsilon_2$ and $\varepsilon_3$. These new studies on correlations between three flow coefficients as well as correlations between higher moments of two different flow coefficients will pave the way to tighten constraints on initial-state models and help to extract precise information on the dynamic evolution of the hot and dense matter created in heavy-ion collisions at the LHC.

\end{abstract}
\ifpreprint
\end{titlepage}                                                                                                                                              
\setcounter{page}{2}
\else
\maketitle
\fi

%=====================INTRODUCTION===============================
\section{Introduction}

One of the fundamental questions in the phenomenology of quantum chromodynamics is what are the properties of matter at extreme densities and temperatures where quarks and gluons are in a state of matter called the quark--gluon plasma (QGP)~\cite{Lee:1978mf, Shuryak:1980tp}. High-energy heavy-ion collisions at the Relativistic Heavy Ion Collider (RHIC) at BNL and the Large Hadron Collider (LHC) at CERN create such a state of strongly interacting matter allowing us to study its properties in the laboratory. Anisotropic flow is a key phenomenon that provides important information about the transport properties of the created QGP matter. Due to large pressure gradients, the anisotropy of the overlapping region between two colliding nuclei causes an anisotropic distribution of the emitted particles in the final state. This anisotropic particle distribution can be quantified by anisotropic flow~\cite{Ollitrault:1992bk, Voloshin:1994mz} which is characterized by the single-particle azimuthal distribution,
\begin{equation}
P(\varphi) = \frac{1}{2\pi} \left[ 1 + 2\, \sum_{n=1}^{\infty} {v_{n} \, \cos n (\varphi - \Psi_{n})} \right].
\end{equation}
Here $\varphi$ is the azimuthal angle of the emitted particle, $v_{n}$ and $\Psi_{n}$ are the $n$-th order flow coefficient and flow symmetry plane, respectively. Both $v_n$ and $\Psi_n$ define the $n$-th order flow-vector as $\overrightarrow{V_{n}} = v_{n}\,e^{in\Psi_{n}}$.
The size and direction of $\overrightarrow{V_{n}}$ related to the initial anisotropy $\overrightarrow{\large \varepsilon_n}$ vector is defined by the moments of the shape of the transverse positions $(r,\phi)$ of the participating nucleons,
\begin{equation}
\overrightarrow{{\huge \varepsilon_n}} = \varepsilon_n \, e^{-in\Phi_n} = -\frac{\left< r^n \, e^{-in\phi} \right>}{\left< r^n \right>},  (n > 1)
\end{equation}
where $\varepsilon_n$ and $\Phi_n$ are the magnitude and orientation of $\overrightarrow{\large \varepsilon_n}$, respectively, and $\left< \, \right>$ stands for the average over all participating nucleons in the initial state.
For lower orders, $n=$ 2 and 3, originally a linear response of $v_n$ to $\varepsilon_n$ was expected, with $v_n = \kappa_n \, \varepsilon_n$~\cite{Kolb:2000sd,Alver:2010gr} where $\kappa_n$ is a parameter that encodes the transport properties of the produced QGP. Later on, it was noticed in models that, already in semi-peripheral collisions, the correlation between the initial $\varepsilon_{2}$ ($\varepsilon_{3}$) and the final-state $v_{2}$ ($v_3$) is not completely linear, with a non-negligible spread in the correlation between $v_n$ and $\varepsilon_n$~\cite{Niemi:2012aj}. Such a nonlinear response of lower-order $v_n$ should be related to the dynamic evolution of the system, but it was briefly investigated in previous studies~\cite{Niemi:2012aj, Noronha-Hostler:2015dbi, Niemi:2015qia}.
For the higher orders, $n\geq4$, $\overrightarrow{V_{n}}$ receives a significant nonlinear hydrodynamic response from $\overrightarrow{\varepsilon}_{2,3}$ in non-central collisions, which was studied in great detail~\cite{Teaney:2012ke, Yan:2015jma, Qian:2016fpi, Aad:2015lwa,Acharya:2017zfg, Acharya:2019uia, Acharya:2020taj, Sirunyan:2019izh}.

One can describe the distribution of final-state anisotropies using a joint probability density function ($p.d.f.$) in terms of $v_{n}$ and $\Psi_{n}$ as $P(v_{m}, v_{n}, ..., \Psi_{m}, \Psi_{n}, ...)$. This is sensitive to the spatial anisotropy $\varepsilon_n$, its event-by-event fluctuations, the correlations between different orders of anisotropy coefficients and initial participant planes $\Phi_n$ carried by $P(\varepsilon_{m}, \varepsilon_{n}, ..., \Phi_{m}, \Phi_{n}, ...)$ and it also reflects the early state dynamics and the transport properties of the QGP. Although ideally one would like to measure $P(v_{m}, v_{n}, ..., \Psi_{m}, \Psi_{n}, ...)$, this is not straightforward to achieve in experiments, but what can be measured are the projections of the full $p.d.f.$ on a finite number of variables~\cite{Jia:2014jca}. Most of these projected distributions could be classified into the following types: (1) $v_n$ fluctuations $P(v_n)$ for both integrated and differential $v_n$ measurements, (2) $\Psi_n$ fluctuations $P(\Psi_{n})$ in different phase space, (3) correlations involving only flow coefficients $P(v_{m}, v_{n}, ...)$, (4) correlations involving only flow symmetry planes $P(\Psi_{m}, \Psi_{n}, ..)$ and (5) mixed correlations carrying both flow coefficients and flow symmetry planes.

The $v_n$ coefficients were measured up to the ninth order with an unprecedented degree of precision~\cite{Acharya:2020taj}. The full $p.d.f.$ of single $v_n$ coefficients $P(v_n)$ was either measured with a Bayesian unfolding procedure~\cite{Aad:2013xma,Sirunyan:2017fts} (for $n=$ 2, 3 and 4) or constructed via the measured moments (for $n=2$)~\cite{Acharya:2018lmh}. It was found that the $P(v_n)$ distribution, which originates from the $p.d.f.$ of initial-state $\varepsilon_n$ distribution $P(\varepsilon_n)$, is described better by an elliptic-power function than a Bessel-Gaussian function~\cite{Acharya:2018lmh}. It was also realized that during the expansion the produced particles might not share a common flow symmetry plane at different transverse momenta, $p_{\rm T}$, and pseudorapidity, $\eta$~\cite{Heinz:2013bua, Gardim:2012im}. These transverse momentum and pseudorapidity dependent flow vectors fluctuate event-by-event, which also breaks the factorization of two-particle correlations $V(p_{\rm T}^{t},p_{\rm T}^{a})$ into the product of flow coefficients $v_{n}(p_{\rm T}^{t}) \cdot v_{n}(p_{\rm T}^{a})$~\cite{Acharya:2017ino, CMS:2013bza, Khachatryan:2015oea}. Such phenomena were predicted by hydrodynamic calculations and are found to be sensitive to the initial-state density fluctuations and/or to the specific shear viscosity of the expanding medium~\cite{Heinz:2013bua, Gardim:2012im, Kozlov:2014hya}. In addition, analyses of correlations between different order flow vectors~\cite{Aad:2014fla, Aad:2015lwa, ALICE:2016kpq} show promise to shed additional light on the initial-state conditions. The correlations between different order symmetry planes were initially investigated in the observable $v_{2n/\Psi_{n}}$~\cite{Andronic:2000cx, Chung:2001qr, Adams:2003zg, ALICE:2011ab}. This was followed by measurements of nonlinear flow modes of higher harmonics by ALICE~\cite{ALICE:2011ab,Acharya:2017zfg, Acharya:2019uia, Acharya:2020taj} as well as event-plane correlations by ATLAS~\cite{Aad:2014fla}. 

The correlation observables involving only anisotropic flow coefficients $v_{m}$ and $v_{n}$ were at first measured with event-shape engineering studies~\cite{Aad:2015lwa} proceeded by investigations using symmetric cumulants~\cite{Bilandzic:2013kga}, defined as $SC(m,n) = \left< v_{n}^{2} \, v_{m}^{2} \right> - \left< v_{n}^{2} \right> \, \left< v_{m}^{2} \right> $. To study such correlations without the dependence on individual flow coefficients, the normalized symmetric cumulant $NSC(m,n)$ was further proposed~\cite{ALICE:2016kpq}. It was found that $NSC(3,2)$, which studies the correlations between $v_2^2$ and $v_3^2$, is very sensitive to the initial conditions and can be used as a good tool to probe initial state $\varepsilon_2^2$ and $\varepsilon_3^2$ correlations. On the other hand, $NSC(4,2)$ and also $NSC$ involving higher order flow coefficients, are sensitive to both initial conditions and the QGP properties. Thus, these $NSC$ measurements have the potential to distinguish between various models of QGP evolution in hydrodynamic and transport models~\cite{Bilandzic:2013kga, Bhalerao:2014xra, Zhou:2015eya, Niemi:2015qia, Zhu:2016puf}. 

It is evident that the study of correlations between various moments of different flow coefficients will deepen our knowledge of the joint $p.d.f.$ for flow magnitudes and angles. However, only correlations involving the second moments of two flow coefficients, $v_n^2$ and $v_m^2$, have been measured utilizing $SC(m,n)$ while the rest have not yet been explored in experiments. In this Letter, an additional step has been made in this direction by using mixed harmonic cumulants ($MHC$)~\cite{Moravcova:2020wnf} to investigate correlations involving more than two different flow coefficients and to study the relationship between higher moments of different flow coefficients in heavy-ion collisions at the LHC. These new measurements establish a milestone for the study of the underlying $p.d.f.$ from $P(v_n)$ to $P(v_n, v_m ...)$, and significantly improve the overall understanding of the initial conditions and the transport properties of the created QGP at the LHC.

%========================METHODS=================================
\section{Observables and methods}

The multiparticle cumulant of mixed harmonics that involves only flow coefficients, named $MHC$, was introduced in Ref.~\cite{Moravcova:2020wnf}. It is defined as an $m$-observable cumulant~\cite{Kubo} in terms of azimuthal angles. By construction, lower order correlations have been subtracted to form genuine multiparticle correlations. Thus, $MHC$ is expected to be insensitive to non-flow effects. This was confirmed in the study of $MHC$ using the HIJING model~\cite{Gyulassy:1994ew}, which does not generate collective flow phenomena~\cite{Moravcova:2020wnf}. For $MHC$ involving only two flow coefficients of second-order, it is identical with the previously defined four-particle symmetric cumulants, i.e. $MHC(v_m^2, v_n^2) = SC(m,n)$. The six-particle cumulant $MHC$ involving $v_2^4$ and $v_3^2$ is
\begin{eqnarray}
MHC(v_2^4, v_3^2) &=& \left<\left<  \cos(2\varphi_1 + 2\varphi_2 + 3\varphi_3 - 2\varphi_4 - 2\varphi_5 - 3\varphi_6) \right>\right> \nonumber\\
&& - 4\,\left<\left<  \cos(2\varphi_1 + 3\varphi_2 - 2\varphi_3 - 3\varphi_4) \right>\right>\,\left<\left<  \cos(2\varphi_1 - 2\varphi_2) \right>\right> \nonumber\\ 
&& - \left<\left<  \cos(2\varphi_1 + 2\varphi_2  - 2\varphi_3 - 2\varphi_4) \right>\right> \,\left<\left<  \cos(3\varphi_1 - 3\varphi_2) \right>\right> \nonumber\\ 
&& + 4\, \left<\left<  \cos(2\varphi_1 - 2\varphi_2) \right>\right>^{2} \, \left<\left<  \cos(3\varphi_1 - 3\varphi_2) \right>\right>.
\label{eq:MHC223223}
\end{eqnarray}
Here the double angular brackets indicate the averaging procedure performed first over all possible combinations of $m$-particle tuples that form the $m$-particle correlation and subsequently the weighted average of all events is calculated with the number of combinations used as an event weight~\cite{Bilandzic:2010jr}. 
In the above expressions, lower order (i.e. two- and four-particle) correlations were removed from the six-particle correlation, which results in a genuine six-particle correlation between $v_{2}^{4}$ and $v_{3}^{2}$. One can rewrite the expression in terms of flow coefficients $v_2$ and $v_3$,
\begin{equation}
MHC(v_2^4, v_3^2) = \left<  v_2^4 \, v_3^2 \right> -  4 \,\left<  v_2^2 \, v_3^2 \right> \, \left<  v_2^2 \right> -  \left<  v_2^4 \right> \, \left<  v_3^2 \right> + 4 \,\left<  v_2^2 \right>^2 \, \left<  v_3^2 \right>.
\end{equation}

Likewise, one can define other six-particle mixed harmonic cumulants that contain $v_2^2$ and $v_{3}^2$ or $v_2^2$, $v_3^2$ and $v_4^2$, in terms of flow coefficients,
\begin{eqnarray}
MHC(v_2^2, v_3^4) &=& \left<  v_2^2 \, v_3^4 \right> -  4 \,\left<  v_2^2 \, v_3^2 \right> \, \left<  v_3^2 \right> -  \left<  v_3^4 \right> \, \left<  v_2^2 \right> + 4 \,\left<  v_2^2 \right> \, \left<  v_3^2 \right>^2,   \\
MHC(v_2^2, v_3^2, v_4^2) &=& \left<  v_2^2 \, v_3^2 \, v_4^2 \right> -  \left<  v_2^2 \, v_3^2 \right> \, \left<  v_4^2 \right> -  \left<  v_2^2 \, v_4^2 \right> \, \left<  v_3^2 \right>  -  \left<  v_3^2 \, v_4^2 \right> \, \left<  v_2^2 \right>   + 2 \,\left<  v_2^2 \right> \, \left<  v_3^2 \right> \, \left<  v_4^2 \right>.
\label{eq:6p_MHC223223}
\end{eqnarray}
Note in general $MHC(v_m^2,v_n^2,v_p^2)$ is different from the so-called higher order symmetric cumulant $SC(m,n,k)$ proposed in~\cite{Mordasini:2019hut}. However, $MHC(v_2^2,v_3^2,v_4^2)$ happens to be the same as $SC(2,3,4)$.

Similarly, the eight-particle mixed harmonic cumulant is defined as an eight-observable cumulant, which can be written in terms of flow coefficients,
\begin{eqnarray}
MHC(v_2^6, v_3^2) &=& \left<  v_2^6 \, v_3^2 \right>  -  9\, \left<  v_2^4 \, v_3^2 \right> \,\left<  v_2^2 \right> - \left<  v_2^6 \right> \, \left< v_3^2 \right> - 9 \, \left<  v_2^4 \right> \, \left<  v_2^2 \, v_3^2 \right>  - 36\, \left<  v_2^2 \right>^3 \, \left< v_3^2 \right>   \nonumber\\ 
& & + 18 \, \left<  v_2^2 \right> \, \left< v_3^2 \right> \, \left<  v_2^4 \right> + 36 \, \left<  v_2^2 \right>^2\,\left<  v_2^2 \, v_3^2 \right>,  \\ 
MHC(v_2^4, v_3^4) &=&  \left< v_2^4 \, v_3^4 \right> - 4 \,  \left< v_2^4 \, v_3^2 \right>  \left< v_3^2 \right>  -  4 \, \left< v_2^2 \, v_3^4 \right> \,  \left<  v_2^2\right> -  \left<  v_2^4 \right> \,  \left< v_3^4 \right> - 8 \,  \left< v_2^2 \, v_3^2 \right>^{2}  \nonumber\\ 
& & - 24\,  \left<  v_2^2 \right>^2 \, \left<  v_3^2 \right>^2 + 4\,  \left< v_2^2 \right>^2 \, \left< v_3^4 \right>  + 4 \, \left<  v_2^4 \right> \,  \left< v_3^2 \right>^2 + 32 \,  \left< v_2^2 \right> \, \left< v_3^2 \right> \,  \left< v_2^2 \, v_3^2\right>,  \\
MHC(v_2^2, v_3^6) &=& \left<  v_2^2 \, v_3^6 \right>  -  9\, \left<  v_2^2 \, v_3^4 \right> \,\left<  v_3^2 \right> - \left<  v_3^6 \right> \, \left< v_2^2 \right>  - 9 \, \left<  v_3^4 \right> \, \left<  v_2^2 \, v_3^2 \right>  - 36\, \left<  v_2^2 \right> \, \left< v_3^2 \right>^3  \nonumber\\ 
& & + 18 \, \left<  v_2^2 \right> \, \left< v_3^2 \right> \, \left< v_3^4 \right> + 36 \, \left<  v_3^2 \right>^2\,\left<  v_2^2 \, v_3^2 \right>.
\end{eqnarray} 

To study genuine multiparticle correlations that are independent of the magnitude of the flow coefficients, the normalized mixed harmonic cumulants $nMHC$ involving two flow coefficients $v_m$ and $v_n$ are constructed according to:
\begin{equation}
nMHC(v_m^k,v_n^l)=\frac{MHC(v_m^k,v_n^l)}{\left\langle v_m^k\right\rangle\left\langle v_n^l\right\rangle},
\label{nMHC}
\end{equation}
Here $m^{k} \neq n^{l}$ to ensure that $nMHC(v_m^k,v_n^l)$ does not contain flow symmetry plane correlations. The expression of Eq.~\ref{nMHC} is also independent of the magnitudes of $v_m$ and $v_n$ and can therefore be used to quantitatively compare genuine correlations between $v_m^k$ and $v_n^l$ determined from experimental data to those determined from the model calculations. Analogously, for $MHC$ involving three flow coefficients without flow symmetry plane correlations, we define the corresponding $nMHC$,
\begin{equation}
nMHC(v_m^k,v_n^l, v_p^q)=\frac{MHC(v_m^k,v_n^l, v_p^q)}{\left\langle v_m^k\right\rangle \, \left\langle v_n^l\right\rangle \, \left\langle v_p^q\right\rangle}.
\label{nMHCH}
\end{equation} 
Here $m^{k} \neq n^{l} \neq p^{q}$, and the sum of any two of $m^{k}$, $n^{l}$ and $p^{q}$ is not equal the third term to avoid flow symmetry plane correlations. Since systematic studies of $v_n$ coefficients were carried out for $n=1 - 9$ in a previous work~\cite{Acharya:2020taj}, this Letter focuses only on the normalized measurements to avoid repeating the earlier discussions on the $v_n$ coefficients themselves. Additionally, multi-particle correlations with sub-event method have been used in the normalizations~\cite{Acharya:2017zfg,Huo:2017nms}, to suppress potential non-flow contamination.

In general, one should be able to construct arbitrary mixed harmonic cumulants to any order. However, due to the limited amount of data available, mixed harmonic cumulants higher than the eighth order will not be examined here. All of the previously mentioned two- and multiparticle azimuthal correlations can be measured by using the latest development of the generic algorithm for multiparticle azimuthal correlations~\cite{Moravcova:2020wnf}.

%===============================DATA SETS========================
\section{Data sets and systematic uncertainty}

This analysis uses data sample from \PbPb\ collisions at $\snn =$ 5.02 TeV recorded with the ALICE detector~\cite{Aamodt:2008zz,Abelev:2014ffa} during the LHC Run 2 (year 2015) data-taking period. Minimum bias events were triggered by a coincidence signal in the two scintillator arrays of the V0 detector, V0A and V0C, which cover the pseudorapidity ranges of $2.8<\eta<5.1$ and $-3.7<\eta<-1.7$, respectively~\cite{Abbas:2013taa}. Only events with a reconstructed primary vertex within $\pm10$ cm from the nominal interaction point along the beam direction were used in this analysis. Removal of background events from, e.g., beam interactions with the residual gas molecules in the beam pipe and pileup events was performed based on the information from the Silicon Pixel Detector (SPD) and the V0 detector. A sample of $55\times10^6$ \PbPb\ collisions, which passed these event selection criteria, were used for the analysis.
 
Charged tracks were reconstructed using the Inner Tracking System (ITS)~\cite{Aamodt:2010aa} and the Time Projection Chamber (TPC)~\cite{Alme:2010ke}. The selected tracks are required to have at least 70 TPC space points (out of a maximum of 159), and the average $\chi^2$ per degree of freedom of the track fit to the TPC space points is required to be lower than two. Additionally, a minimum of two hits are required in the ITS to improve the momentum resolution. A selection requiring the pseudorapidity to be within $|\eta|< 0.8$ is applied. Tracks with a transverse momentum $\pt < 0.2$~GeV/$c$ or $\pt > 5.0$~GeV/$c$ were rejected due to the magnetic field and to reduce the contribution from jets, respectively~\cite{Adam:2016izf}. In addition, a criterion on the maximum distance of closest approach of the track to the collision point of less than $2$~cm in the longitudinal direction and less than a $\pt$-dependent selection in the transverse direction, ranging from $0.2$~cm at $\pt=0.2$ GeV/$c$ to $0.016$~cm at $\pt=5.0$ GeV/$c$, was applied. This results in a residual contamination from secondary particles from weak decays and from interactions in the detector material of 1--3\%, which is negligible in the final systematic uncertainty. These selection criteria result in a transverse momentum dependent efficiency of track reconstruction of about 80\%.

Numerous potential sources of systematic uncertainty were investigated in the analysis, including variations of the event and track selection and the uncertainty associated with possible remaining non-flow effects. These are the azimuthal angle correlations not associated with the common flow symmetry planes, including contributions from jets, resonance decays, and are estimated using the HIJING model and found to be negligible for all of the presented observables. The variation of the results with the choice of collision centrality was calculated by alternatively using the SPD to estimate the event multiplicity and was found to contribute less than 5\% for all observables. Results with opposite polarities of the magnetic field within the ALICE detector and with narrowing the nominal $\pm$10 cm range of the reconstructed vertex along the beam direction from the center of the ALICE detector to 9, 8, and 7 cm showed a difference of 0--5.4\% compared to results with the default selection criteria. The contribution from pileup events was investigated by varying the selections on the correlations between multiplicities from the V0 and SPD, and was found to be negligible. The sensitivity to the track selection criteria was explored by varying the number of TPC space points and by comparing the results to those obtained with tracks with different requirements on hits in the ITS. The effect of varying the number of TPC space points from 70 to 80 and 90 resulted in a negligible systematic uncertainty. Using different track requirements led to a difference with respect to the default selection criteria of less than 4.3\% except for $nMHC(v_{2}^{2}, v_{3}^{2}, v_{4}^{2})$ where it was about 16\%. 
The systematic uncertainty evaluated for each above-mentioned source found to be statistically significant according to the recommendation in~\cite{Barlow:2002yb} were added in quadrature to obtain the measurements' total systematic uncertainty.

%================================RESULTS==========================

\section{Results}

\begin{figure}[htbp!]
\begin{center}
\includegraphics[width=0.7\linewidth]{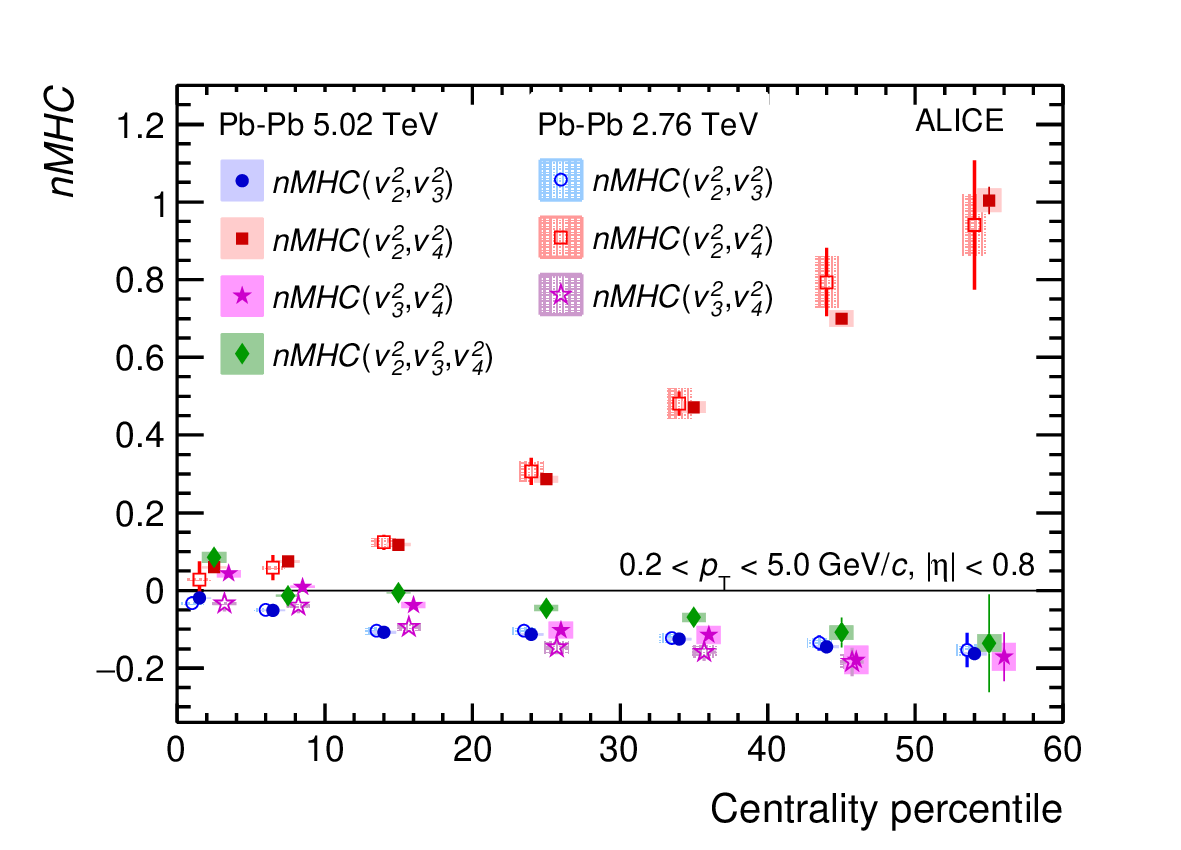}
\caption{Centrality dependence of $nMHC(v_{2}^{2}, v_{3}^{2})$,  $nMHC(v_{2}^{2}, v_{4}^{2})$, $nMHC(v_{3}^{2}, v_{4}^{2})$ and $nMHC(v_{2}^{2}, v_{3}^{2}, v_{4}^{2})$ in Pb--Pb collisions at $\sqrt{s_{\rm NN}} = 5.02$~TeV, shown by the solid markers. The statistical (systematic) errors are shown with vertical bars (filled boxes). Comparisons to the previous measurements at 2.76 TeV from Refs.~\cite{ALICE:2016kpq,Acharya:2017gsw}, shown by the open markers, are also presented. Data points are shifted for visibility.}
\label{fig:2k_3l_4p}
\end{center}
\end{figure}

The centrality dependence of mixed harmonic cumulants with two and three flow coefficients are measured in Pb--Pb collisions at $\sqrt{s_{\rm NN}} = 5.02$~TeV. The results of $nMHC(v_{2}^{2}, v_{3}^{2})$, $nMHC(v_{2}^{2}, v_{4}^{2})$, $nMHC(v_{3}^{2}, v_{4}^{2})$, and $nMHC(v_{2}^{2}, v_{3}^{2}, v_{4}^{2})$ are presented in Fig.~\ref{fig:2k_3l_4p} by blue solid circles, red solid squares, magenta solid stars and green diamonds, respectively. Positive values of $nMHC(v_{2}^{2}, v_{4}^{2})$ and negative values of $nMHC(v_{2}^{2}, v_{3}^{2})$ are observed for all centralities, which means that $v_{2}^2$ and $v_{4}^2$ are correlated while $v_{2}^2$ and $v_{3}^2$ are anti-correlated. This indicates that finding $v_2$ larger than $\left< v_2 \right>$ in an event enhances the probability of finding $v_4$ larger than $\left< v_4 \right>$ and $v_3$ smaller than $\left< v_3 \right>$ in that event. For $nMHC(v_{3}^{2}, v_{4}^{2})$, a similar centrality dependence as for $nMHC(v_{2}^{2}, v_{3}^{2})$ is seen for centralities above 20--30\% where the nonlinear hydrodynamic response of $\overrightarrow{V_{4}}$ plays a significant role~\cite{Acharya:2017zfg, Acharya:2020taj}. These new measurements are compared in Fig.~\ref{fig:2k_3l_4p} to the previously published results at $\sqrt{s_{\rm NN}} = 2.76$~TeV, which were named $SC(m,n)$ in Ref.~\cite{ALICE:2016kpq,Acharya:2017gsw}, shown with open markers. The results of $nMHC(v_{2}^{2}, v_{3}^{2})$,  $nMHC(v_{2}^{2}, v_{4}^{2})$ are compatible within uncertainties at the two different energies, which indicates a weak dependence on the collision energy of these two observables. However, there are differences for $nMHC(v_{3}^{2}, v_{4}^{2})$ between the two studied energies, which increase towards central collisions. In particular, the measurement at 5.02 TeV changes sign from negative to positive in central collisions, while it remains negative at the lower energy. A similar study of multiparticle cumulants in the most central collisions was investigated in great detail in Ref.~\cite{Aaboud:2019sma}, where a significant effect from centrality fluctuations was found in Pb--Pb collisions at 5.02 TeV. Moreover, the amplitude of the centrality fluctuations depends on how the centrality was determined. 

Besides the measurements of correlations between two flow coefficients, the new measurement of correlations between three flow coefficients, $nMHC(v_{2}^{2}, v_{3}^{2}, v_{4}^{2})$, is shown with green diamonds in Fig.~\ref{fig:2k_3l_4p}. 
As introduced above, $nMHC(v_{2}^{2}, v_{3}^{2}, v_{4}^{2})$ is identical to $SC(2,3,4)$, which has been recently measured at a lower energy~\cite{Acharya:2021afp}. By construction, the lower order few-particle correlations have been subtracted from the higher order correlations in the $nMHC$. Thus, it is expected that $nMHC(v_{2}^{2}, v_{3}^{2}, v_{4}^{2})$ should be consistent with zero, if the correlations between three flow coefficients are purely driven by the correlations between two flow coefficients. It is seen in Fig.~\ref{fig:2k_3l_4p} that the result of $nMHC(v_{2}^{2}, v_{3}^{2}, v_{4}^{2})$ is located between the $nMHC(v_{2}^{2}, v_{3}^{2})$, $nMHC(v_{3}^{2}, v_{4}^{2})$ and $nMHC(v_{2}^{2}, v_{4}^{2})$ for the centrality classes under study. More specifically, it is positive and closer to $nMHC(v_{2}^{2}, v_{4}^{2})$ and $nMHC(v_{3}^{2}, v_{4}^{2})$ in the most central collisions, then it changes sign to negative and shows a similar centrality dependence to $nMHC(v_{2}^{2}, v_{3}^{2})$ and $nMHC(v_{3}^{2}, v_{4}^{2})$ in non-central collisions. The non-zero result of $nMHC(v_{2}^{2}, v_{3}^{2}, v_{4}^{2})$ for the presented centrality range shows the existence of genuine correlations between three flow coefficients and thus brings new information toward determining $P(v_{n}, v_{m}, ...)$ that cannot be obtained from measurements of correlations between two flow coefficients.

\begin{figure}[htbp!]
\begin{center}
\includegraphics[width=0.45\linewidth]{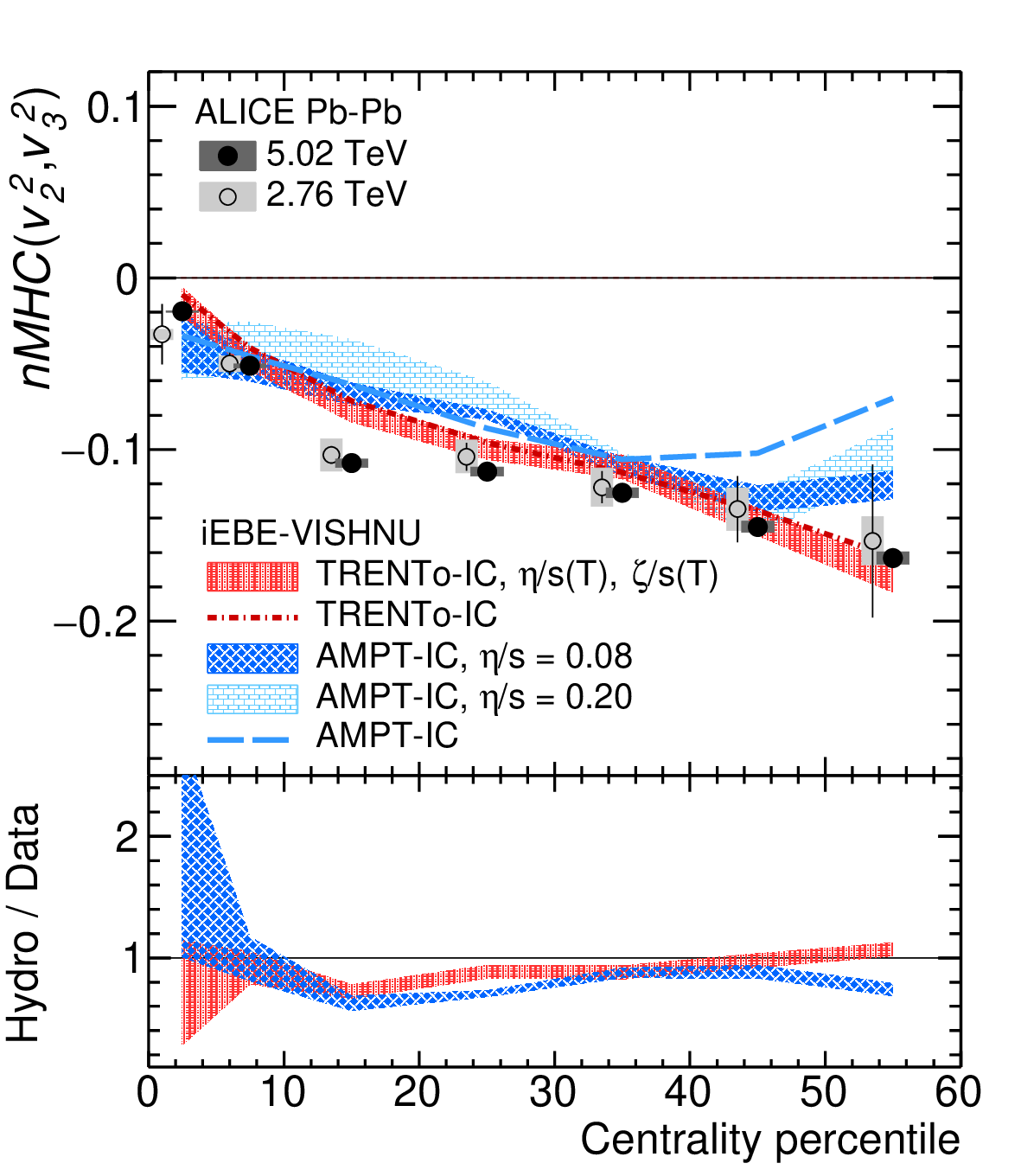}
\includegraphics[width=0.45\linewidth]{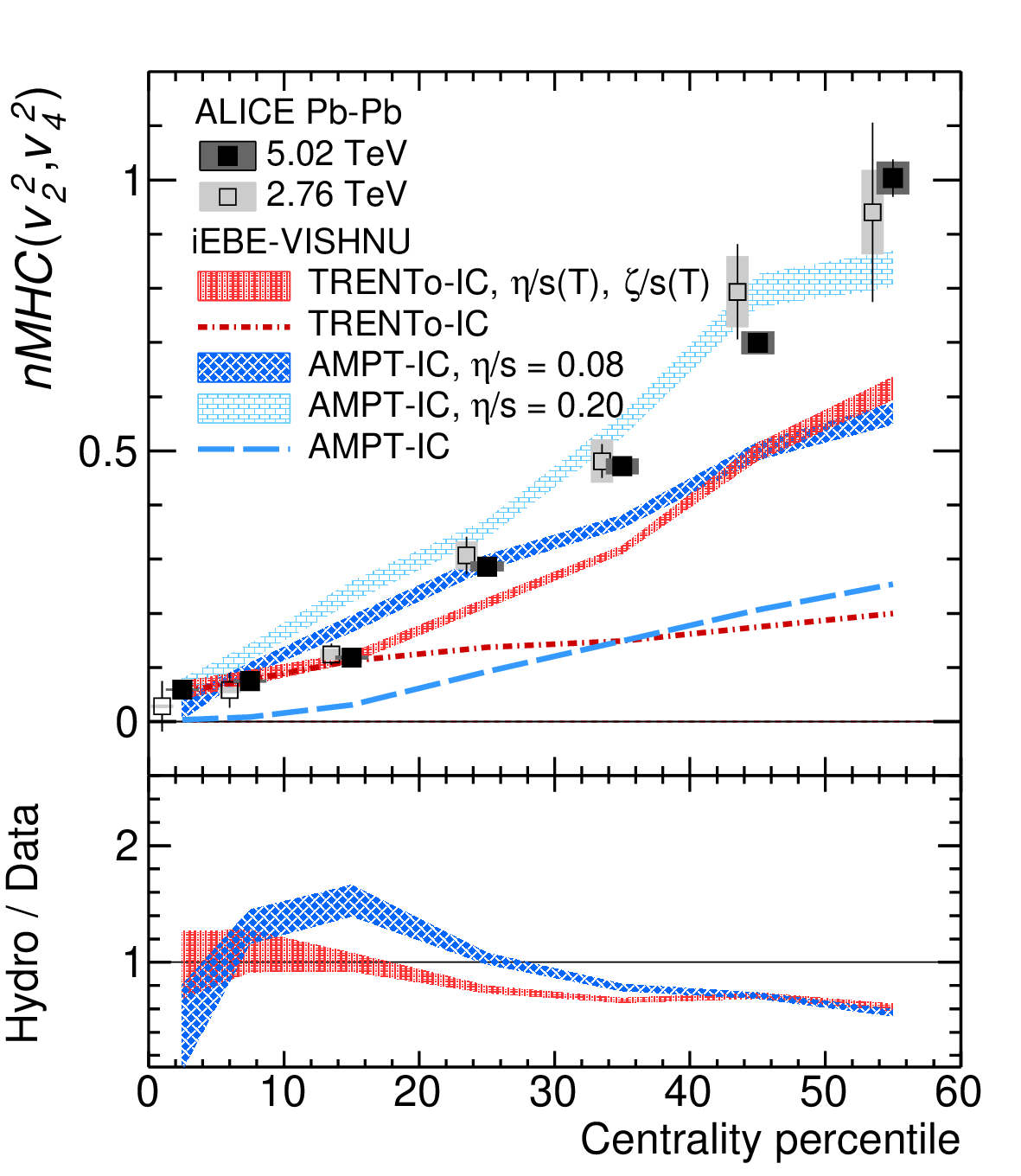}
\caption{Centrality dependence of $nMHC(v_{2}^{2}, v_{3}^{2})$ and $nMHC(v_{2}^{2}, v_{4}^{2})$ in \PbPb\ collisions at $\sqrt{s_{\rm NN}} = 5.02$~TeV (solid markers) and 2.76 TeV (open markers). Statistical uncertainties are shown as vertical bars and systematic uncertainties as filled boxes. The iEBE-VISHNU calculations~\cite{Zhao:2017yhj} for Pb--Pb collisions at 5.02 TeV with TRENTo initial conditions (red shadowed bands) and AMPT initial conditions (blue shadowed bands) are presented, together with the corresponding initial state calculations $nMHC(\varepsilon_{2}^{2}, \varepsilon_{3}^{2})$, $nMHC(\varepsilon_{2}^{2}, \varepsilon_{4}^{2})$ from the TRENTo (red dot-dash lines) and AMPT model (blue long-dash lines). The same marker (line) styles and colors are used in later figures. Data points are shifted for visibility.}
\label{fig2} 
\end{center}
\end{figure}

\begin{figure}[htbp!]
\begin{center}
\includegraphics[width=0.45\linewidth]{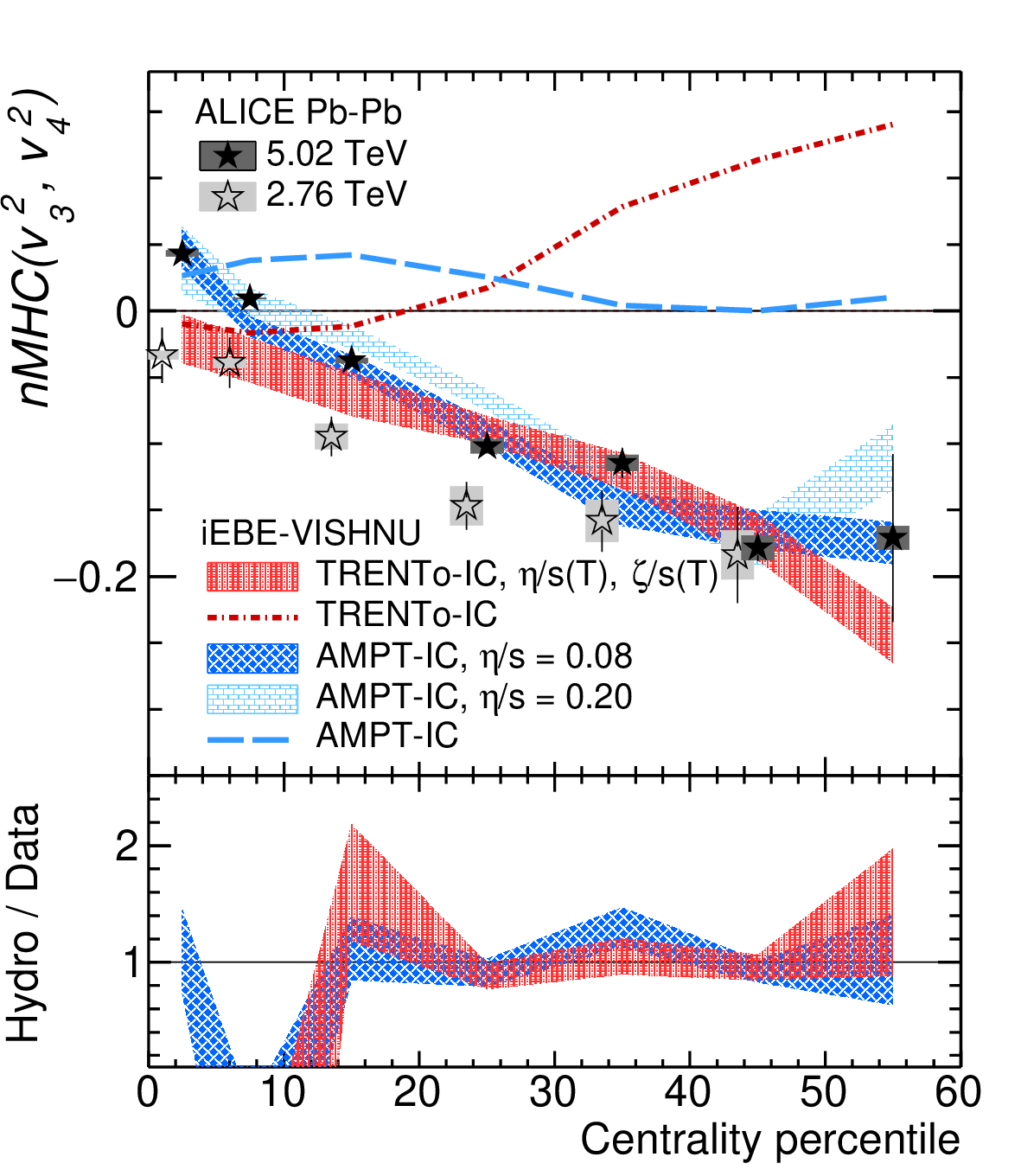}
\includegraphics[width=0.45\linewidth]{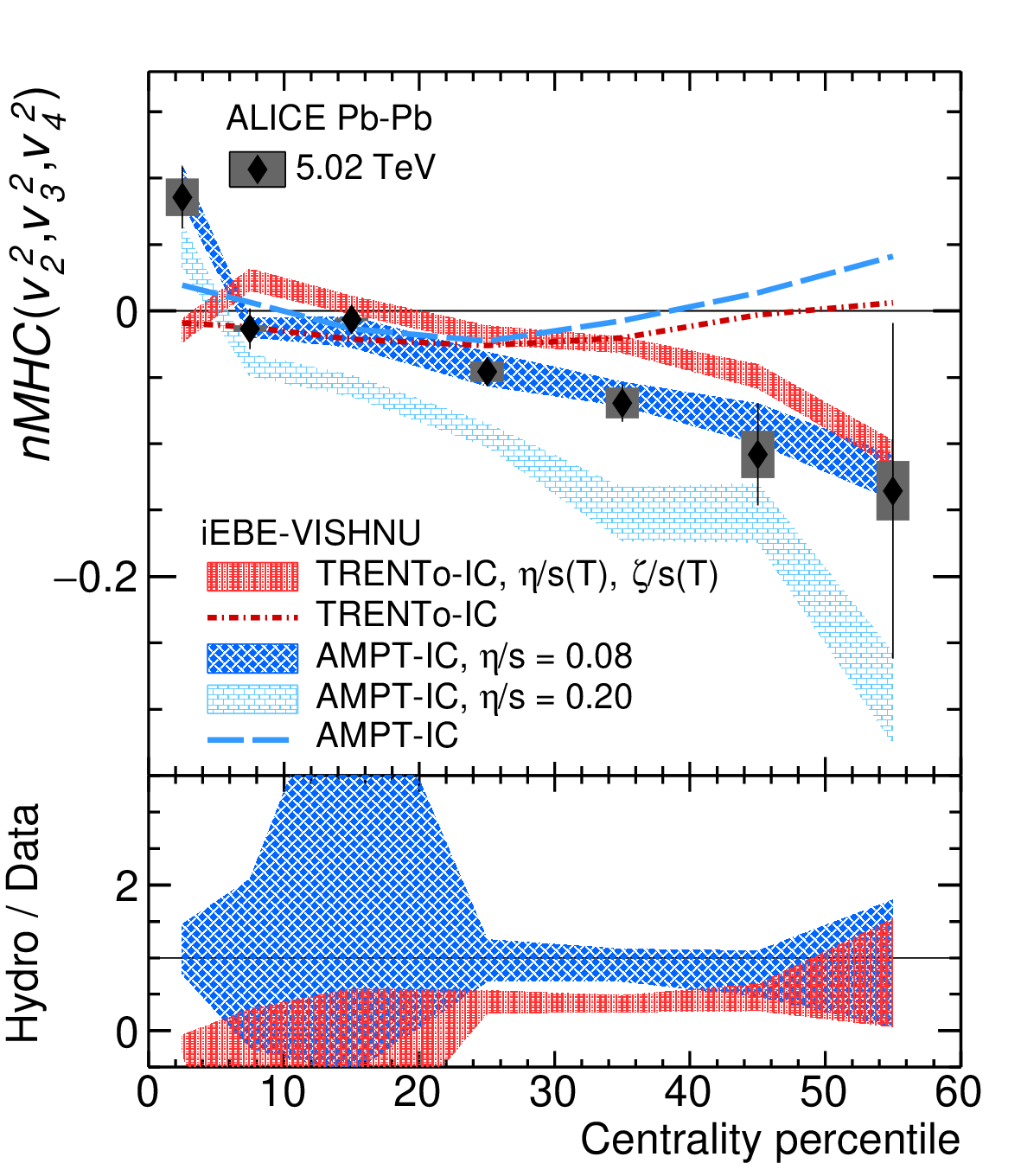}
\caption{Centrality dependence of $nMHC(v_{3}^{2}, v_{4}^{2})$ and $nMHC(v_{2}^{2}, v_{3}^{2}, v_{4}^{2})$ in \PbPb\ collisions at $\sqrt{s_{\rm NN}} = 5.02$~TeV (solid markers) and 2.76 TeV (open markers). Statistical uncertainties are shown as vertical bars and systematic uncertainties as filled boxes. Data points are shifted for visibility.}
\label{fig3} 
\end{center}
\end{figure}

In order to gain more information on the initial conditions and transport properties of the created QGP at the LHC, the results are compared with those from hydrodynamic model calculations. Results from the hybrid iEBE-VISHNU model with TRENTo initial conditions with specific shear viscosity $\eta/s(T)$ and bulk viscosity $\zeta/s(T)$ extracted from the best fit of a Bayesian analysis~\cite{Bernhard:2019bmu} as well as calculations with AMPT-initial conditions with $\eta/s = 0.08$ and no bulk viscosity~\cite{Zhao:2017yhj} are compared to the data. Both calculations can quantitatively describe the flow coefficients from inclusive and identified hadrons~\cite{Zhao:2017yhj,Acharya:2018zuq} and also provide a reasonable description of more complicated flow observables, e.g., nonlinear modes of higher-order flow~\cite{Acharya:2020taj,Acharya:2019uia}. Besides these two calculations, the iEBE-VISHNU model with AMPT-initial conditions with $\eta/s = 0.20$ and no bulk viscosity is also used. This model does not describe the particle spectra nor the flow coefficients and thus should not be compared with the experimental data. In the remaining text, the hydrodynamic calculations using AMPT initial conditions and $\eta/s = 0.08$ will be refereed to as "AMPT calculations". However, the comparison of hydrodynamic calculation from the same initial state model but with different $\eta/s$ values can be very useful to study the sensitivity of various $nMHC$ to the $\eta/s$ of the QGP.

Comparisons of the measured $nMHC(v_{2}^{2}, v_{3}^{2})$ and $nMHC(v_{2}^{2}, v_{4}^{2})$ to hydrodynamic calculations are shown in Fig.~\ref{fig2}. 
In general, the hydrodynamic calculations with both AMPT and TRENTo initial conditions, shown as blue and red shadowed bands, respectively, predict qualitatively the centrality dependence of $nMHC$. In addition, as $v_2$ and $v_3$ are linearly correlated with the initial $\varepsilon_{2}$ and $\varepsilon_{3}$ in central and semi-central collisions, compatible results of the final-state $nMHC(v_{2}^{2}, v_{3}^{2})$ calculations and the $nMHC(\varepsilon_{2}^{2}, \varepsilon_{3}^{2})$ calculations from the initial-state models are expected~\cite{Moravcova:2020wnf}. This is indeed shown by the shaded areas and the dashed lines in Fig.~\ref{fig2} (left). In the same figure, there is also no difference between the calculations using AMPT-initial conditions with different $\eta/s$ values. This suggests that for the presented centrality ranges, $v_2^2$ ($v_3^2$) is linearly correlated with the initial $\varepsilon_2^2$ ($\varepsilon_3^2$). Thus, the $nMHC(v_{2}^{2}, v_{3}^{2})$ measurements shown in Fig.~\ref{fig2} (left) can be used to directly constrain the correlations between the initial anisotropy coefficients $\varepsilon_2^2$ and $\varepsilon_3^2$ without much consideration of the exact value of the transport coefficients in the hydrodynamic models. For $nMHC(v_{2}^{2}, v_{4}^{2})$ results shown in Fig.~\ref{fig2} (right), both calculations underestimate the data; the TRENTo calculation fits the data better in central collisions, while the AMPT calculation works slightly better for centralities above 20\%. The initial-state calculations of $nMHC(\varepsilon_{2}^{2}, \varepsilon_{4}^{2})$ are significantly lower than the final-state $nMHC(v_{2}^{2}, v_{4}^{2})$ calculations, which suggests that the correlation between $v_{2}^{2}$ and $v_{4}^{2}$ is not driven solely by the initial correlation between $\varepsilon_{2}^{2}$ and $\varepsilon_{4}^{2}$, but it is mainly developed at later stages of the system's dynamic evolution, especially the nonlinear response contribution to $v_4$.

Figure~\ref{fig3} (left) compares hydrodynamic calculations with the $nMHC(v_{3}^{2}, v_{4}^{2})$ measurement. In general, both models generate the same trend of centrality dependence as is seen in data. Notably, the AMPT calculations also predict the sign change in central collisions, while the TRENTo calculations remain negative for the entire centrality range. It has also been seen in Ref.~\cite{Zhao:2017yhj} that the AMPT calculations always predict a positive correlation in the most central collisions at 2.76 and 5.02 TeV, while the TRENTo calculations are always negative at both collision energies. Although the hydrodynamic calculations of $nMHC(v_{3}^{2}, v_{4}^{2})$ from AMPT and TRENTo initial conditions are almost compatible for non-central collisions, the initial correlations between $\varepsilon_{3}^{2}$ and $\varepsilon_{4}^2$, quantified by $nMHC(\varepsilon_{3}^{2}, \varepsilon_{4}^{2})$, are utterly different from the two initial-state models, and are far away from the final-state $nMHC(v_{3}^{2}, v_{4}^{2})$ calculations. It could be attributed to a significant nonlinear hydrodynamic response in $v_{4}$ from $\varepsilon_2^2$. This nonlinear contribution is strongly anti-correlated with $\varepsilon_{3}^2$ and plays a dominant role in the final $nMHC(v_{3}^{2}, v_{4}^{2})$ results for non-central collisions. On the other hand, in the same centrality region, the linear response of $v_{4}$ to $\varepsilon_4$ is rather weak~\cite{Acharya:2017zfg}, and the contributions from correlations between the initial $\varepsilon_{3}^{2}$ and $\varepsilon_{4}^2$ in the final $nMHC(v_{3}^{2}, v_{4}^{2})$ appear to be minor.

To extend the discussion from correlations of two flow coefficients to three flow coefficients, the measurement of $nMHC(v_{2}^{2}, v_{3}^{2}, v_{4}^{2})$ and its comparison to hydrodynamic calculations with both AMPT and TRENTo initial conditions are presented in Fig.~\ref{fig3} (right). In general, the agreement between the initial $nMHC(\varepsilon_{2}^{2}, \varepsilon_{3}^{2}, \varepsilon_{4}^{2})$ correlations and the final-state $nMHC(v_{2}^{2}, v_{3}^{2}, v_{4}^{2})$ calculations worsens as the collision centrality becomes more peripheral, which can be expected due to the increasing contribution from the nonlinear hydrodynamic response in $v_4$. Figure~\ref{fig3} (right) also shows clearly that the calculation with AMPT initial conditions and $\eta/s=0.08$ describes the data reasonably well. The calculation with $\eta/s = 0.20$ is two times larger than the one with $\eta/s=0.08$. Such a difference is more significant compared to what has been seen in the correlations between two harmonics, where no obvious difference is observed for $nMHC(v_{2}^{2}, v_{3}^{2})$ and $nMHC(v_{3}^{2}, v_{4}^{2})$ and only a relatively small difference is seen for $nMHC(v_{2}^{2}, v_{4}^{2})$. 
This demonstrates the novelty of the new correlations between three flow coefficients constraining the transport properties of the QGP. However, despite the fact that the hydrodynamic calculations using TRENTo initial conditions are consistent with the measured $nMHC(v_{2}^{2}, v_{3}^{2})$ and $nMHC(v_{3}^{2}, v_{4}^{2})$, and also provide a reasonable description of the $nMHC(v_{2}^{2}, v_{4}^{2})$ measurement, they significantly underestimate the data by roughly a factor of two. Considering an apparent discrepancy between the data and TRENTo calculations, there is little doubt that the hydrodynamic framework and its corresponding parameters can be better tuned in a future Bayesian analysis if this new $nMHC(v_{2}^{2}, v_{3}^{2}, v_{4}^{2})$ measurement is used as an input. The first measurement of correlations between three harmonics provides additional independent constraints on the theoretical models beyond those provided by the correlations of two harmonics that have been studied before.

\begin{figure}[htbp!] 
\begin{center}
\includegraphics[width=0.7\linewidth]{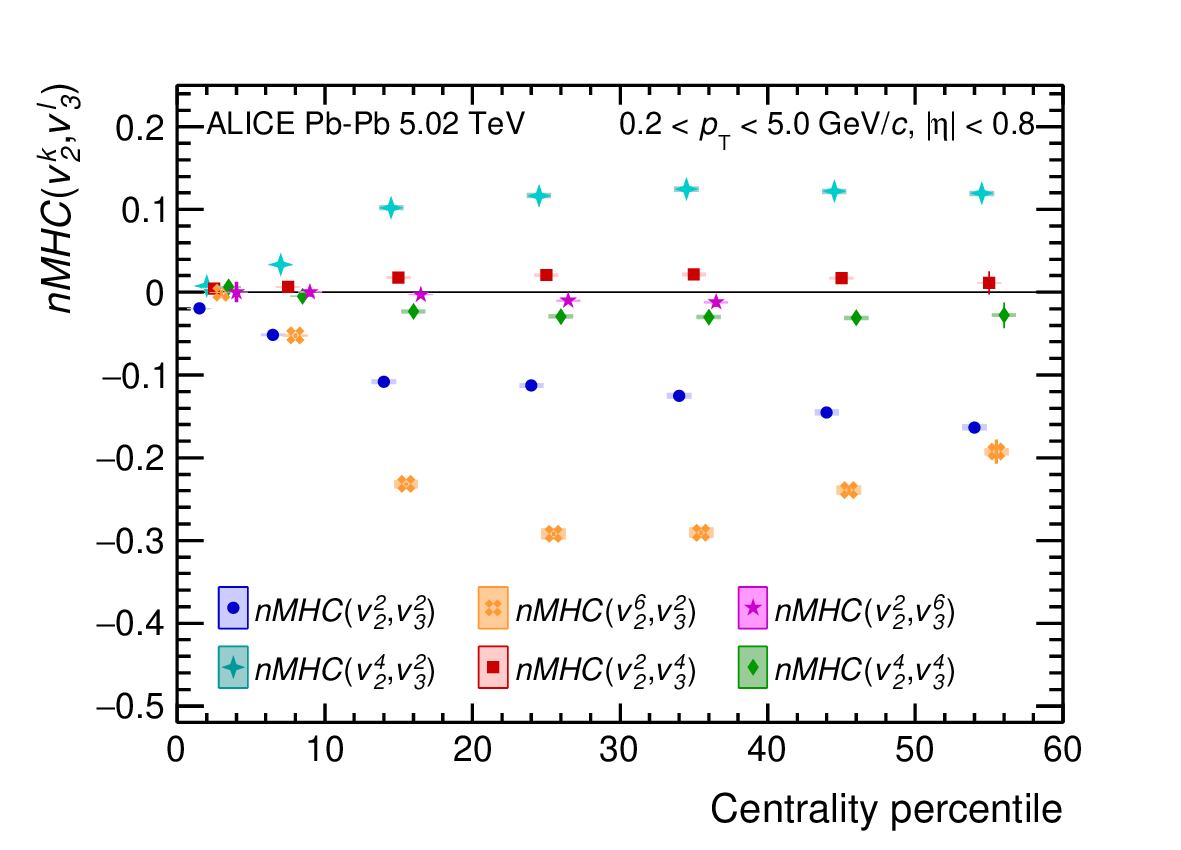}
\caption{Centrality dependence of $nMHC$ for \PbPb\ collisions at $\sqrt{s_{\rm NN}} = 5.02$~TeV. 
	Statistical uncertainties are shown as vertical bars and systematic uncertainties as filled boxes. Data points are shifted for visibility.}
\label{fig4} 
\end{center}
\end{figure}

With the recently proposed observable $nMHC$, one can study not only the correlations between two or three different flow coefficients, in terms of their second moments, but also the correlations between the $k^{th}$ order moment of $v_{m}$ and the $l^{th}$ order moment of $v_n$ where $k \geq 2$, $l \geq 2$. It is particularly interesting to study the correlations between various moments of $v_2$ and $v_3$ because in central and semi-central collisions, both $v_2$ and $v_3$ are linearly correlated to their corresponding initial eccentricities $\varepsilon_2$ and $\varepsilon_3$~\cite{Noronha-Hostler:2015dbi, Niemi:2015qia}. Thus the measurement of $nMHC(v_{2}^{k}, v_{3}^{l})$ in central and semi-central collisions might provide a direct constraint on the initial correlation between $\left<\varepsilon_2^k\right>$ and $\left<\varepsilon_3^{l} \right>$. This information is extremely important for the understanding of the initial conditions of heavy-ion collisions but it has never been measured before. Conversely, the potential nonlinearity of $v_2$ and $v_3$, more pronounced in peripheral collisions, strongly depends on the dynamical evolution of the created QGP. The study of $nMHC(v_{2}^{k}, v_{3}^{l})$ will enable a new way to study this effect on $v_2$ and $v_3$. 

\begin{figure}[htbp!]
\begin{center}
\includegraphics[width=0.45\linewidth]{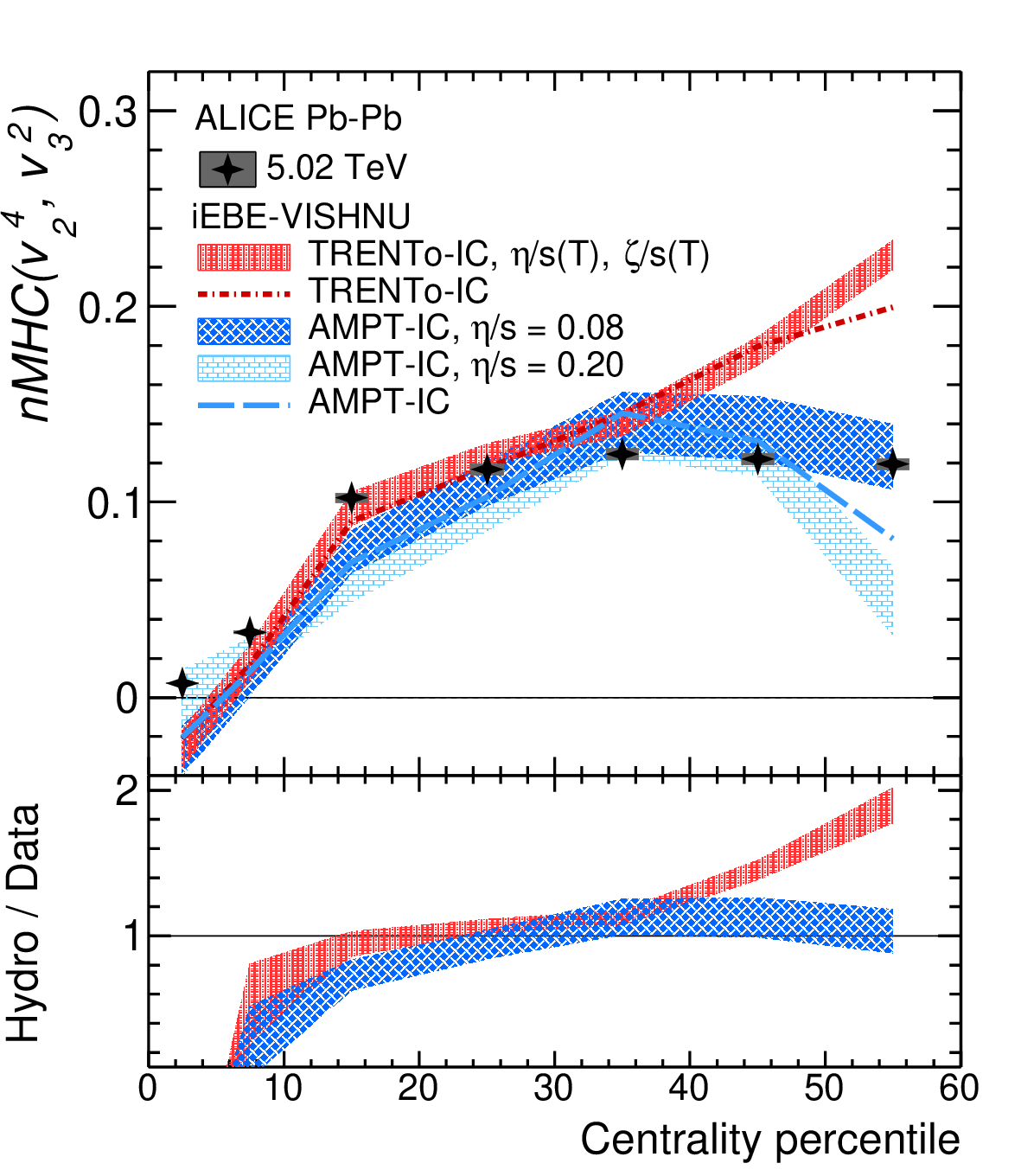}
\includegraphics[width=0.45\linewidth]{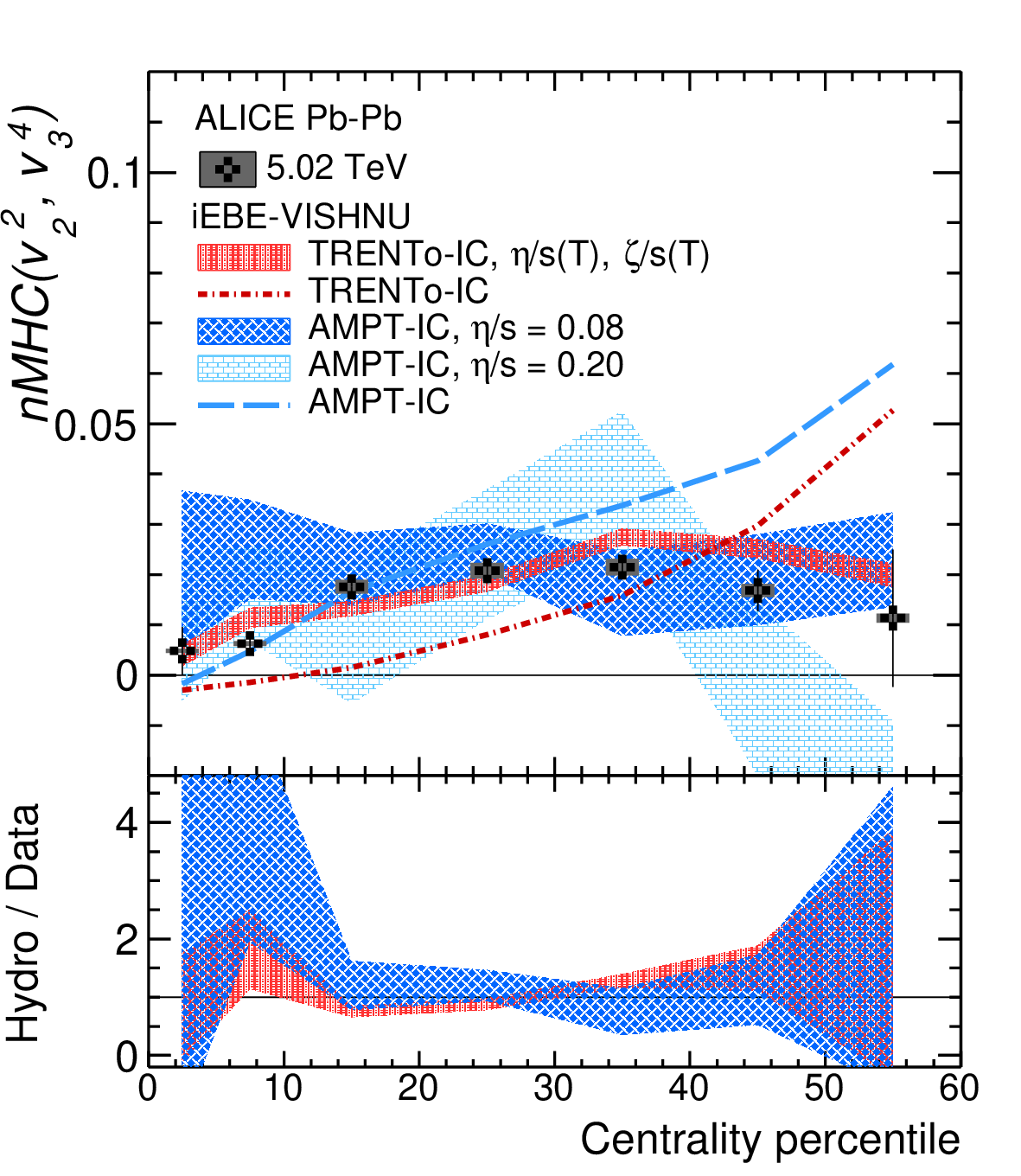}
\caption{Centrality dependence of $nMHC(v_{2}^{4}, v_{3}^{2})$ and $nMHC(v_{2}^{2}, v_{3}^{4})$ in \PbPb\ collision at $\sqrt{s_{\rm NN}} = 5.02$~TeV. Statistical uncertainties are shown as vertical bars and systematic uncertainties as filled boxes. }
\label{fig5} 
\end{center}
\end{figure}

The first measurements of $nMHC(v_{2}^{k}, v_{3}^{l})$ are presented in Fig.~\ref{fig4}. In addition to the negative value of $nMHC(v_{2}^{2}, v_{3}^{2})$, which has been studied before, positive correlations are observed for both six-particle mixed harmonic cumulants $nMHC(v_{2}^{4}, v_{3}^{2})$ and $nMHC(v_{2}^{2}, v_{3}^{4})$. The eight-particle mixed harmonic cumulants $nMHC(v_{2}^{6}, v_{3}^{2})$, $nMHC(v_{2}^{4}, v_{3}^{4})$ and $nMHC(v_{2}^{2}, v_{3}^{6})$ are all negative. Such characteristic negative, positive and negative signs of four-, six-, and eight-particle mixed harmonic cumulants, respectively, are very similar to the previously measured pattern for two-, four-, six,- and eight-particle single harmonic cumulants in Pb--Pb collisions~\cite{Acharya:2018lmh}, which show positive, negative, positive, and negative signs, respectively. 
These findings agree qualitatively with the initial-state predictions based on the MC-Glauber~\cite{Moravcova:2020wnf}, AMPT, and TRENTo models~\cite{Zhao:2017yhj}. It should be pointed out that the measured negative $nMHC(v_{2}^{2}, v_{3}^{2})$ shown above could only confirm the negative correlations of $(v_2^{2}, v_3^2)$, while the results presented in Fig.~\ref{fig4} illustrate further the positive correlations of $(v_2^{4}, v_3^{2})$ and $(v_2^{2}, v_3^{4})$ as well as the negative correlations of $(v_2^{6}, v_3^{2})$, $(v_2^{4}, v_3^{4})$ and $(v_2^{2}, v_3^{6})$. Moreover, one can see the following hierarchy, $ |nMHC(v_{2}^{6}, v_{3}^{2})| > |nMHC(v_{2}^{2}, v_{3}^{2})| \geq |nMHC(v_{2}^{4}, v_{3}^{2})| > |nMHC(v_{2}^{4}, v_{3}^{4})| \approx |nMHC(v_{2}^{2}, v_{3}^{4})| > |nMHC(v_{2}^{2}, v_{3}^{6})|.$ 
This agrees qualitatively with the predictions based on initial-state models~\cite{Moravcova:2020wnf,Zhao:2017yhj}. Furthermore, the calculations based on the HIJING model~\cite{Gyulassy:1994ew}, which does not generate anisotropic flow in the created system, are consistent with zero~\cite{Moravcova:2020wnf}, and thus do not reproduce the characteristic signs of the multiparticle mixed harmonic cumulants observed in experiments. Future studies with other non-flow models, i.e. PYTHIA~\cite{Sjostrand:2014zea, Bierlich:2020naj}, could confirm if the aforementioned characteristic signs of the multiparticle mixed harmonic cumulants can be regarded as a flow signature and thus could be used for searching for collective flow in small collision systems like pp or pA collisions~\cite{Loizides:2016tew, Song:2017wtw, Nagle:2018nvi}.

\begin{figure}[htbp!]
\begin{center}
\includegraphics[width=0.45\linewidth]{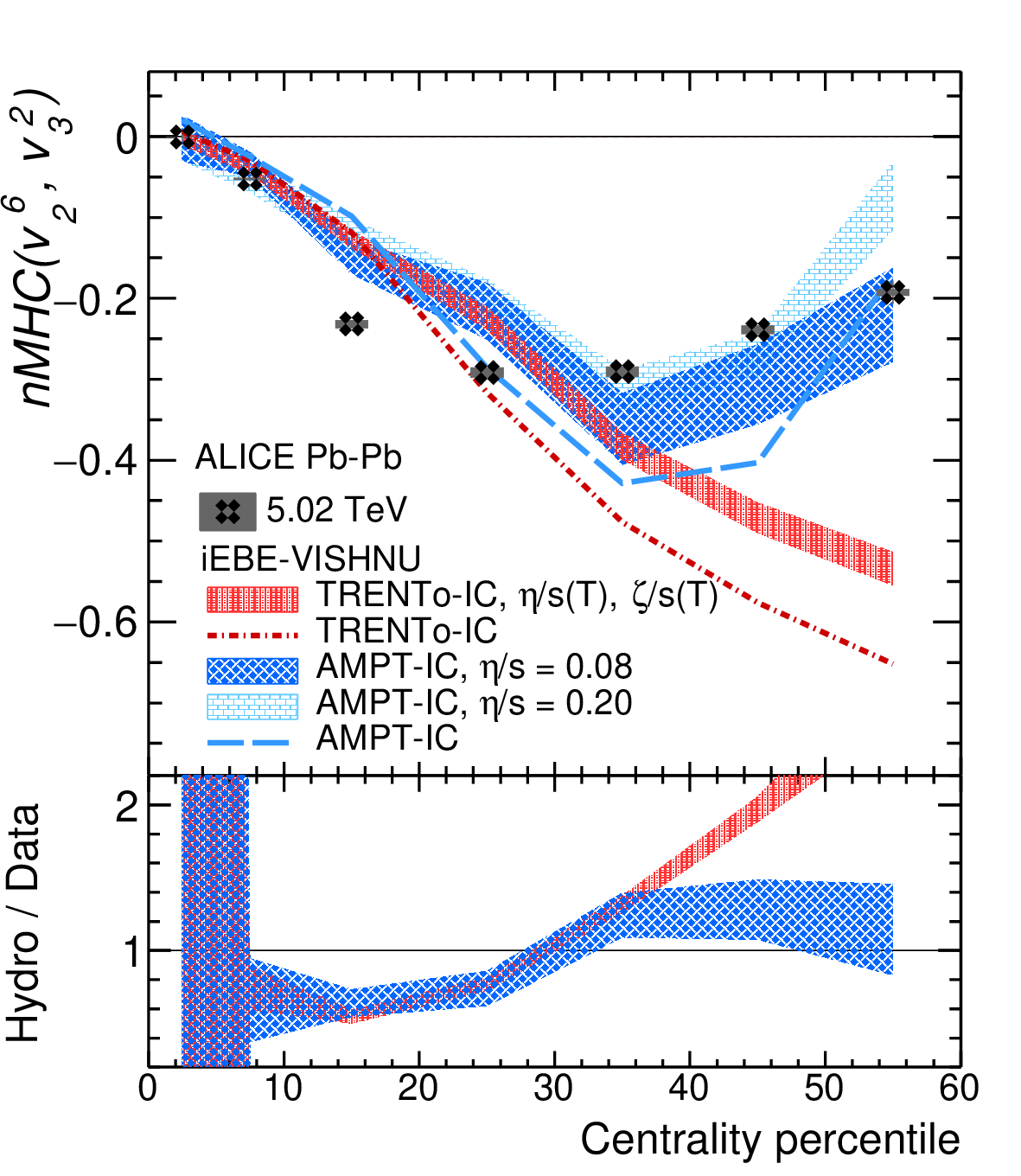}
\includegraphics[width=0.45\linewidth]{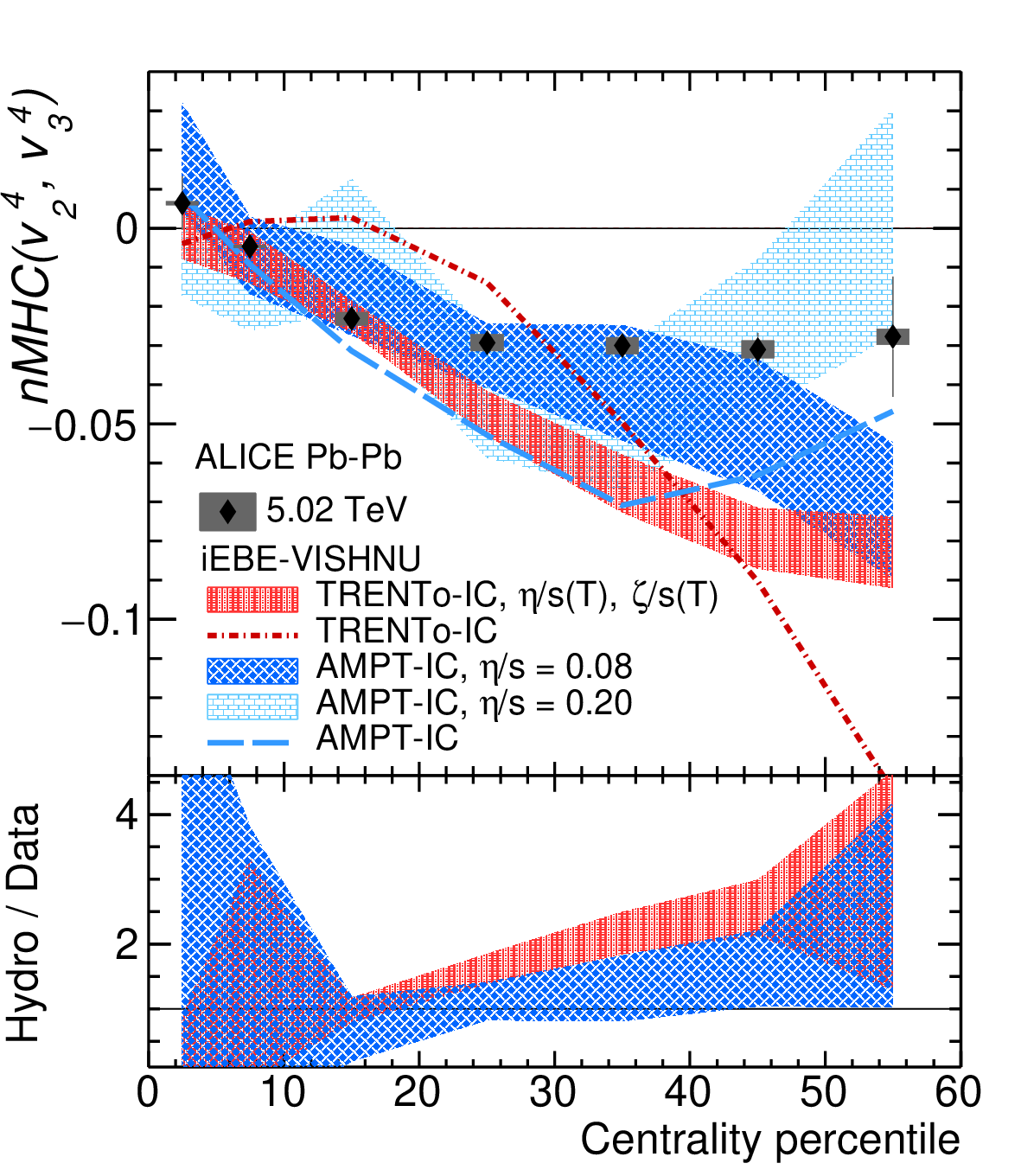}
\caption{Centrality dependence of $nMHC(v_{2}^{6}, v_{3}^{2})$ (left) and $nMHC(v_{2}^{4}, v_{3}^{4})$ (right) in \PbPb\ collisions at $\sqrt{s_{\rm NN}} = 5.02$~TeV. Statistical uncertainties are shown as vertical bars and systematic uncertainties as filled boxes.  }
\label{fig6} 
\end{center}
\end{figure}

As mentioned above, for non-peripheral collisions, both $v_2$ and $v_3$ are expected to be linearly correlated with the initial eccentricity $\varepsilon_2$ and $\varepsilon_3$. Thus, the final-state result of $nMHC(v_{2}^{k}, v_{3}^{l})$ could reflect the initial correlation between $\varepsilon_2^k$ and $\varepsilon_3^l$. This behavior is observed in the case of $nMHC(v_{2}^{2}, v_{3}^{2})$, where good agreement with $nMHC(\varepsilon_{2}^{2}, \varepsilon_{3}^{2})$ was found. Moving to higher moments of $v_2$ and/or $v_3$, one can further probe the nonlinearity of $v_2$ ($v_3$) to $\varepsilon_{2}$ ($\varepsilon_3$) by seeing if the agreement between the initial and final-state correlations persists, because of the better sensitivity of higher moments to the nonlinear hydrodynamic response.
Figure~\ref{fig5} presents the comparison of data with iEBE-VISHNU calculations with AMPT and TRENTo initial conditions. This figure shows that both calculations describe the measured $nMHC(v_{2}^{4}, v_{3}^{2})$ fairly well for central and semi-central collisions. The calculations with AMPT initial conditions work better for more peripheral collisions. At the same time, consistent results are observed between $nMHC(\varepsilon_{2}^{4}, \varepsilon_{3}^{2})$ and $nMHC(v_{2}^{4}, v_{3}^{2})$, independent of whether the AMPT or TRENTo initial-state model are used. For AMPT calculations, there is no difference between the results using $\eta/s = 0.08$ or 0.20, which confirms that the precision measurement of $nMHC(v_{2}^{4}, v_{3}^{2})$ can offer an additional approach to constrain initial-state models. However, the situation is different in the case of $nMHC(v_{2}^{2}, v_{3}^{4})$, shown in Fig.~\ref{fig5} (right). The hydrodynamic calculations with both AMPT and TRENTo initial conditions are compatible with the measurement within the considerable uncertainty, but there is an apparent discrepancy between $nMHC(v_{2}^{2}, v_{3}^{4})$ and $nMHC(\varepsilon_{2}^{2}, \varepsilon_{3}^{4})$. This does not agree with the naive expectation of both $v_2$ and $v_3$ being linearly correlated with their respective initial eccentricities $\varepsilon_2$ and $\varepsilon_3$, which might be because, generally, the linearity of $v_{3}$ to $\varepsilon_3$ is worse than that of $v_2$ to $\varepsilon_2$ as shown by hydrodynamic calculations~\cite{Niemi:2012aj}. When one examines higher-order moments, the linear response of $v_2^4$ remains and thus $nMHC(\varepsilon_{2}^{4}, \varepsilon_{3}^{2}) = nMHC(v_{2}^{4}, v_{3}^{2})$ is observed. However, the nonlinearity of $v_3^4$ becomes non-negligible in non-peripheral collisions, which creates the discrepancy between the initial $nMHC(\varepsilon_{2}^{2}, \varepsilon_{3}^{4})$ and final state $nMHC(v_{2}^{2}, v_{3}^{4})$ correlations.

\begin{figure}[htbp!]
\begin{center}
\includegraphics[width=0.45\linewidth]{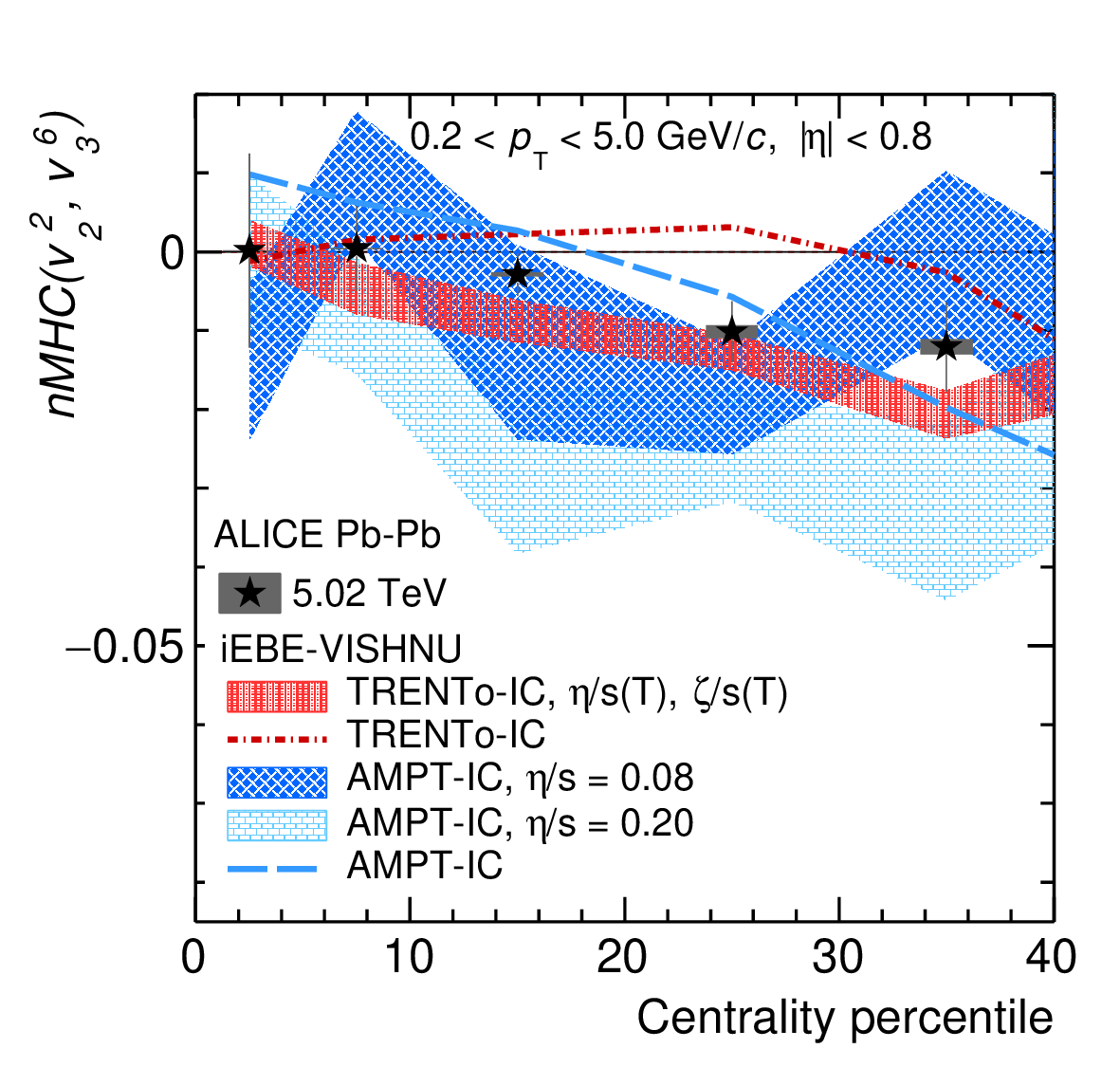}
\caption{Centrality dependence of $nMHC(v_{2}^{2}, v_{3}^{6})$ in \PbPb\ collisions at $\sqrt{s_{\rm NN}} = 5.02$~TeV. Statistical uncertainties are shown as vertical bars and systematic uncertainties as filled boxes. }
\label{fig7} 
\end{center}
\end{figure}

This hypothesis is further confirmed in Figs.~\ref{fig6} and ~\ref{fig7} where eight-particle cumulants are reported, which involve even higher moments of $v_2$ and/or $v_3$. Firstly, hydrodynamic calculations with AMPT initial conditions quantitatively predict the new measurements of $nMHC(v_{2}^{6}, v_{3}^{2})$, $nMHC(v_{2}^{4}, v_{3}^{4})$ and $nMHC(v_{2}^{2}, v_{3}^{6})$, while the TRENTo calculations show compatible results except for $nMHC(v_{2}^{4}, v_{3}^{4})$, where the calculations overestimate the data by a factor of two. Secondly, although in hydrodynamic calculations with two different initial conditions there is a similar centrality dependence of $nMHC(\varepsilon_{2}^{6}, \varepsilon_{3}^{2})$ and $nMHC(v_{2}^{6}, v_{3}^{2})$, a clear difference between the two is observed already in semi-central collisions. This difference becomes much larger when fourth and sixth order moments of $v_3$ are involved, with no obvious agreement between the initial and final-state calculations; this is especially shown by the TRENTo calculations in Figs.~\ref{fig6} (right) and~\ref{fig7}. It is expected that the effect of the nonlinearity of $v_2$ and $v_3$ will be enhanced when studying $nMHC(v_{2}^{k}, v_{3}^{l})$ with higher moments (e.g. $k, l \geq 4$). The resulting $nMHC(v_{2}^{k}, v_{3}^{l})$ in the final state, instead of being determined solely by the initial correlations of $nMHC(\varepsilon_{2}^{k}, \varepsilon_{3}^{l})$, receive non-negligible contributions from the nonlinearities of higher moments of $v_2$ and $v_3$, developed during the dynamic evolution of the system. Thus, the new measurements of $nMHC(v_{2}^{k}, v_{3}^{l})$ presented in this Letter provide direct access to the initial correlations between $\varepsilon_2^k$ and $\varepsilon_3^{l}$ when lower moments of $v_2$ and $v_3$ are involved, while enabling a new possibility to study the nonlinearities of $v_2$ and $v_3$ when higher moments are involved.

%================================SUMMARY==========================
\section{Summary}
The normalized mixed harmonic cumulants $nMHC$ between two and three flow coefficients as well as between higher moments of two flow coefficients, were measured in $\sqrt{s_{\rm NN}} = 5.02$~TeV Pb--Pb collisions with ALICE. It is found that $nMHC(v_{2}^{2}, v_{4}^{2})$  is positive, while $nMHC(v_{2}^{2}, v_{3}^{2})$ and $nMHC(v_{3}^{2}, v_{4}^{2})$ are negative. In addition, the first measurement of three harmonic correlations $nMHC(v_{2}^{2}, v_{3}^{2}, v_{4}^{2})$ is closer to $nMHC(v_{2}^{2}, v_{4}^{2})$ and $nMHC(v_{3}^{2}, v_{4}^{2})$ in central collisions, and then becomes closer to $nMHC(v_{2}^{2}, v_{3}^{2})$ and $nMHC(v_{3}^{2}, v_{4}^{2})$ for more peripheral collisions. These measurements compared with iEBE-VISHNU hydrodynamic calculations using AMPT and TRENTo initial conditions exhibit different sensitivities to the initial conditions and the specific shear viscosity of the QGP. Thus the measurements presented in this Letter can be used to more tightly constrain theoretical models. Furthermore, the correlations between higher moments of $v_2$ and $v_3$ were investigated for the first time. The four-, six- and eight-particle mixed cumulants of $nMHC$ show the characteristic signature of negative, positive, and negative signs, respectively, similar to the multiparticle cumulants of single harmonics. The comparison with hydrodynamic calculations reveals that the correlations involving higher-order moments could significantly enhance the contributions that arise from nonlinearities of $v_2$ and $v_3$ to the initial eccentricity $\varepsilon_2$, triangularity $\varepsilon_3$, respectively. Such contributions mainly develop during the expansion of the system and reflect the time evolution of the shear and bulk viscosities of the QGP. These new measurements of correlations between different moments of two and three flow coefficients, together with comparisons to state-of-the-art hydrodynamic calculations, provide further information on the initial conditions and considerably tighten the constraints on the evolution of the QGP created in heavy-ion collisions at the LHC.

%%%%%%%%%%%%%%%%%%%%%%%%%%%%%%%%%%%%%%%%%%%%%%%%%%
\newenvironment{acknowledgement}{\relax}{\relax}
%%%%%%%%%%%%%%%%%%%%%%%%%%%%%%%%%%%%%%%%%%%%%%%%%%
\begin{acknowledgement}
\section*{Acknowledgments}

% Version: 2021-02-10

The ALICE Collaboration would like to thank all its engineers and technicians for their invaluable contributions to the construction of the experiment and the CERN accelerator teams for the outstanding performance of the LHC complex.
The ALICE Collaboration gratefully acknowledges the resources and support provided by all Grid centres and the Worldwide LHC Computing Grid (WLCG) collaboration.
The ALICE Collaboration acknowledges the following funding agencies for their support in building and running the ALICE detector:
A. I. Alikhanyan National Science Laboratory (Yerevan Physics Institute) Foundation (ANSL), State Committee of Science and World Federation of Scientists (WFS), Armenia;
Austrian Academy of Sciences, Austrian Science Fund (FWF): [M 2467-N36] and Nationalstiftung f\"{u}r Forschung, Technologie und Entwicklung, Austria;
Ministry of Communications and High Technologies, National Nuclear Research Center, Azerbaijan;
Conselho Nacional de Desenvolvimento Cient\'{\i}fico e Tecnol\'{o}gico (CNPq), Financiadora de Estudos e Projetos (Finep), Funda\c{c}\~{a}o de Amparo \`{a} Pesquisa do Estado de S\~{a}o Paulo (FAPESP) and Universidade Federal do Rio Grande do Sul (UFRGS), Brazil;
Ministry of Education of China (MOEC) , Ministry of Science \& Technology of China (MSTC) and National Natural Science Foundation of China (NSFC), China;
Ministry of Science and Education and Croatian Science Foundation, Croatia;
Centro de Aplicaciones Tecnol\'{o}gicas y Desarrollo Nuclear (CEADEN), Cubaenerg\'{\i}a, Cuba;
Ministry of Education, Youth and Sports of the Czech Republic, Czech Republic;
The Danish Council for Independent Research | Natural Sciences, the VILLUM FONDEN and Danish National Research Foundation (DNRF), Denmark;
Helsinki Institute of Physics (HIP), Finland;
Commissariat \`{a} l'Energie Atomique (CEA) and Institut National de Physique Nucl\'{e}aire et de Physique des Particules (IN2P3) and Centre National de la Recherche Scientifique (CNRS), France;
Bundesministerium f\"{u}r Bildung und Forschung (BMBF) and GSI Helmholtzzentrum f\"{u}r Schwerionenforschung GmbH, Germany;
General Secretariat for Research and Technology, Ministry of Education, Research and Religions, Greece;
National Research, Development and Innovation Office, Hungary;
Department of Atomic Energy Government of India (DAE), Department of Science and Technology, Government of India (DST), University Grants Commission, Government of India (UGC) and Council of Scientific and Industrial Research (CSIR), India;
Indonesian Institute of Science, Indonesia;
Istituto Nazionale di Fisica Nucleare (INFN), Italy;
Institute for Innovative Science and Technology , Nagasaki Institute of Applied Science (IIST), Japanese Ministry of Education, Culture, Sports, Science and Technology (MEXT) and Japan Society for the Promotion of Science (JSPS) KAKENHI, Japan;
Consejo Nacional de Ciencia (CONACYT) y Tecnolog\'{i}a, through Fondo de Cooperaci\'{o}n Internacional en Ciencia y Tecnolog\'{i}a (FONCICYT) and Direcci\'{o}n General de Asuntos del Personal Academico (DGAPA), Mexico;
Nederlandse Organisatie voor Wetenschappelijk Onderzoek (NWO), Netherlands;
The Research Council of Norway, Norway;
Commission on Science and Technology for Sustainable Development in the South (COMSATS), Pakistan;
Pontificia Universidad Cat\'{o}lica del Per\'{u}, Peru;
Ministry of Education and Science, National Science Centre and WUT ID-UB, Poland;
Korea Institute of Science and Technology Information and National Research Foundation of Korea (NRF), Republic of Korea;
Ministry of Education and Scientific Research, Institute of Atomic Physics and Ministry of Research and Innovation and Institute of Atomic Physics, Romania;
Joint Institute for Nuclear Research (JINR), Ministry of Education and Science of the Russian Federation, National Research Centre Kurchatov Institute, Russian Science Foundation and Russian Foundation for Basic Research, Russia;
Ministry of Education, Science, Research and Sport of the Slovak Republic, Slovakia;
National Research Foundation of South Africa, South Africa;
Swedish Research Council (VR) and Knut \& Alice Wallenberg Foundation (KAW), Sweden;
European Organization for Nuclear Research, Switzerland;
Suranaree University of Technology (SUT), National Science and Technology Development Agency (NSDTA) and Office of the Higher Education Commission under NRU project of Thailand, Thailand;
Turkish Atomic Energy Agency (TAEK), Turkey;
National Academy of  Sciences of Ukraine, Ukraine;
Science and Technology Facilities Council (STFC), United Kingdom;
National Science Foundation of the United States of America (NSF) and United States Department of Energy, Office of Nuclear Physics (DOE NP), United States of America.

\end{acknowledgement}

%%%%%%%%%%%%%%%%%%%%%%%%%%%%%%%%%%%%%%%%%%%%%%%%%%
\bibliographystyle{utphys}
\bibliography{bibliography}
%%%%%%%%%%%%%%%%%%%%%%%%%%%%%%%%%%%%%%%%%%%%%%%%%%
\newpage
\appendix
%%%%%%%%%%%%%%%%%%%%%%%%%%%%%%%%%%%%%%%%%%%%%%%%%%
\section{The ALICE Collaboration}
\label{app:collab}
%%%%%%%%%%%%%%%%%%%%%%%%%%%%%%%%%%%%%%%%%%%%%%%%%%

\small
\begin{flushleft} 

S.~Acharya$^{\rm 142}$, 
D.~Adamov\'{a}$^{\rm 97}$, 
A.~Adler$^{\rm 75}$, 
J.~Adolfsson$^{\rm 82}$, 
G.~Aglieri Rinella$^{\rm 35}$, 
M.~Agnello$^{\rm 31}$, 
N.~Agrawal$^{\rm 55}$, 
Z.~Ahammed$^{\rm 142}$, 
S.~Ahmad$^{\rm 16}$, 
S.U.~Ahn$^{\rm 77}$
% ...

\bigskip

\bigskip 

\textbf{\Large Affiliation Notes}

\bigskip 

% ...
$^{\rm V}$ Also at: Institute of Theoretical Physics, University of Wroclaw, Poland\\

\bigskip

\bigskip 

\textbf{\Large Collaboration Institutes}

\bigskip 

% ...
$^{16}$ Department of Physics, Aligarh Muslim University, Aligarh, India\\
% ...
$^{31}$ Dipartimento DISAT del Politecnico and Sezione INFN, Turin, Italy\\
% ...
$^{35}$ European Organization for Nuclear Research (CERN), Geneva, Switzerland\\
% ...
$^{55}$ INFN, Sezione di Bologna, Bologna, Italy\\
% ...
$^{75}$ Johann-Wolfgang-Goethe Universit\"{a}t Frankfurt Institut f\"{u}r Informatik, Fachbereich Informatik und Mathematik, Frankfurt, Germany\\
% ...
$^{77}$ Korea Institute of Science and Technology Information, Daejeon, Republic of Korea\\
% ...
$^{82}$ Lund University Department of Physics, Division of Particle Physics, Lund, Sweden\\
% ...
$^{97}$ Nuclear Physics Institute of the Czech Academy of Sciences, \v{R}e\v{z} u Prahy, Czech Republic\\
% ...
$^{142}$ Variable Energy Cyclotron Centre, Homi Bhabha National Institute, Kolkata, India\\
% ...

\end{flushleft}

%============================

\end{document}